\documentclass[aps,prb,twocolumn,groupedaddress]{revtex4}

\usepackage{graphicx}
\usepackage{dcolumn}
\usepackage{bm}

\begin{document}



\title{Atomic level micromagnetic model of recording media switching at elevated temperatures}


\author{J. I. Mercer, M. L. Plumer and J. P. Whitehead }
\affiliation{Department of Physics and Physical Oceanography, Memorial University, St. John's, Newfoundland  A1B 3X7  Canada}
\author{J. van Ek}
\affiliation{Western Digital Corporation, San Jose, CA 94588 USA }

\date{\today}

\begin{abstract}
An atomic level micromagnetic model of granular recording media is developed and applied to examine external field-induced grain switching at elevated temperatures which captures non-uniform reversal modes.  The results are compared with traditional methods which employ the Landau-Lifshitz-Gilbert equations based on uniformly magnetized grains with assigned intrinsic temperature profiles for $M(T)$ and $K(T)$.  Using nominal parameters corresponding to high-anisotropy FePt-type media envisioned for Energy Assisted Magnetic Recording, our results demonstrate that atomic-level reversal  slightly reduces the field required to switch grains at elevated temperatures, but results in larger fluctuations, when compared to a uniformly magnetized grain model.  
\end{abstract}


\maketitle
Micromagnetics, based on solutions of the dynamic Landau-Lifshitz-Gilbert (LLG) equation, has proven to be an important tool for the evaluation and refinement of new concepts which promise to yield decreases in magnetic recording bit sizes.\cite{plumer00} An important application of micromagnetics is the simulation of recording-media grain magnetization reversal under the influence of the applied transducer write field. The traditional LLG model based on the assumption of weakly coupled uniformly magnetized grains (with dimensions on the order of 10 nm) at zero temperature (where M is constant) provides a reasonable description of the materials and recording processes used in current media.  Although media grains are composed of thousands of atomic spins ${\bf S}_i$ which can react individially to an applied field, strong intra-grain exchange coupling is assumed to force all atomic spin vectors to remain parallel to each other. Proposed technologies such as those involving energy assisted magnetization reversal (EAMR) though the application of a spatially and temporally confined burst of heat or  microwave transverse field \cite{rottmayer06,zhu08,plumer10} have forced a re-evaluation of the traditional LLG model of recording media.   The addition of thermal fluctuations, for example, leads to a reduction in overall grain magnetization ${\bf M} = \sum_i {\bf S}_i$  through statistical processes which increasing randomize the {\it directions} of the individual atomic spins ${\bf S}_i$ so that the thermal average $\langle M \rangle \rightarrow 0$ at temperatures above the intrinsic ferromagnetic Curie temperature $T_c$.   As recently demonstrated, grain magnetization reversal can involve complicated non-linear dynamic nucleation and non-uniform reversal mechanisms at the atomic-spin level for temperatures which are an appreciable fraction of $T_c$.\cite{rohart10,bunce10,leblanc10} 

   There have been a number of different types of models proposed recently which address various aspects associated with higher temperature effects in magnetic grain reversal.  An analytic model based on uniformly magnetized grains which demonstrates broadening in the switching field distribution and transition width due to thermal fluctuations has been proposed.\cite{wang10}   Grain-level micromagnetic simulations based on the LLG equations have also been performed which account for the thermal-fluctuation induced reduction in the intrinsic grain moment and uniaxial anisotropy (K) through the introduction of temperature dependent profiles for M(T) and K(T) into the simulation.\cite{li08,torabi09,plumer10b} The results of these studies illustrate the utility of the thermal EAMR idea by integrating these concepts into a model of the recording process. A more sophisticated approach to account for  the spin structure internal to grains at elevated temperatures through the introduction of longitudinal degrees of freedom within the general framework of micromagnetics (called the LLB (Landau-Lifshitz-Bloch) method) has also been examined.\cite{bunce10,chubykalo06,kazantseva08} This formalism captures some important features of high-temperature grain reversal but remains phenomenological in nature and exhibits deviations from an exact atomic spin-level simulations close to the Curie temperature,\cite{kazantseva08} which is the region of most interest for thermal EAMR recording. 
   
\begin{figure}[h]
\includegraphics[width=0.4\textwidth]{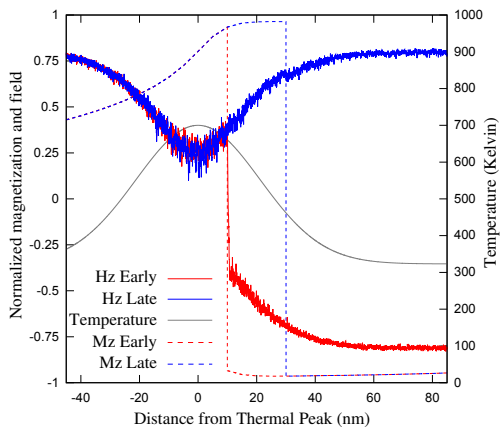}
\caption{\label{one} (color online) Grain switching under the influence of a thermal spot and write field.  Shown are the center-track write field profiles offset 15 nm (early: broken red line) and 30 nm (late: broken blueline) from the thermal spot (grey line) centered at 0 nm.  Maximum write field has a value of about 21 kOe and the maximum temperature is 700 K.  Also shown are grain $M_z/M_s$ switch profiles for 50 runs corresponding to early and late application of the write field (red and blue lines).}
\end{figure}

  In the present work, we report on the results of dynamic atomic-level LLG simulations of recording-media grain magnetization reversal under the influence of the applied transducer write field at elevated temperatures. To examine the importance of accounting for the impact of a reversal field on individual atomic spins in a grain, we examine the recording of transitions in high anisotropy media under the influence of a moving write field and thermal `spot' as envisioned for thermal EAMR.  Individual grain reversal is seen to occur as a highly non-linear process due to the combined effects of the thermal fluctuations and Zeeman energy, $ - {\bf H} \cdot {\bf S}_i$, acting on individual spin vectors.  This level of detail is not accounted for in previously proposed recording models of thermal-fluctuation-induced grain reversal. The present results are compared with the above-mentioned appoach which integrate $M(T)$ and $K(T)$ profiles into the LLG equations.\cite{li08,torabi09}    
  
  Solutions to the standard LLG equation with the stochastic thermal field were obtained using the Euler integration method with damping constant $\alpha = 0.1$, a time step of $0.001$ ps, and with rotations updated using a quaternions representation.\cite{meloche10} Magnetic parameters were based on generic single-layer FePt-type media with a saturation moment $M_s = 1000$  emu/cc, inter-grain exchange was omitted and uniaxial anisotropy was $K=5 \times 10^7$ erg/cc, giving $H_K = 100$ kOe.  Note that our simulations require a zero temperature estimate of the anisotropy, which is nearly a factor of two larger than its value at 300 K.\cite{thiele02} We also tested our method on ultra-high $H_K = 250$ kOe media (by reducing $M_s$ to 400 emu/cc).   Magnetostatic interactions were not included in this preliminary study and this approximation will not have a significant impact on our conclusions since $H_{\rm demag} \sim$ 10 kOe is much smaller than the anisotropy field.  The write field was taken from finite element simulations of a generic design for perpendicular recording giving a maximum field of about 21 kOe in the media. The thermal spot was assumed to have a Gaussian spatial distribution with a full width at half maximum of 50 nm and temperatures from 323.15 K to a peak of 700 K. Each grain was assigned a new temperature at every time step based on the location of the center of the grain as it passed under the thermal spot. This is illustrated in Fig. 1 which shows the center-track profiles of the field and temperature.
  
  An atomic level model of grains was constructed from $10 \times 10 \times 20$ cells  where each cell has dimensions $a \times a \times a$ with $a=0.5 ~$ nm and represents an atomic moment ${\bf S}_i$. The anisotropy parameter K was the same as above and the inter-spin (intra-grain) exchange parameter was $A$=1.35 $\mu$erg/cm, giving $H_{\rm ex}^{\rm intra}$ = 540 kOe.  Simulations of $M(T)$ using these parameters yields $T_c \simeq 700$ K. Grain level models represent each grain with a single magnetic moment ${\bf M}$ having a magnitude which is made to vary with temperature.\cite{torabi09} Both models are simulated using the standard LLG equation as described above.

  Switching distributions corresponding to 50 separate stochastic simulations for a single grain ($M_s = 1000$ emu/cc) subject to these fields, as the grain moves down center track, are shown in Fig. 1.  The speed of the media relative to the head was assumed to be $25$ nm/ns. Two cases were considered. For the `early' runs, the distance between the trailing edge of the write pole when the field reversal occurs and maximum of the thermal spot was 15 nm.  In the `late' simulations, this distance was increased to 30 nm.  In the early case, all of the 50 runs resulted in successful switching.  In the late case, none of the runs produced a switch.  The magnetization switching that occured for the 50 runs as a function of the position of the pole tip are also shown in Fig. 1.  Similar switching behavior was found in the case of $M_s$ = 400 emu/cc (not shown). These results clearly demonstrate that switching occurs only when there is adequate overlap between regions of strong write field and high enough temperature.  They also emphasize the statistical nature of the switching process.
  
\begin{figure}[h]
\includegraphics[width=0.5\textwidth]{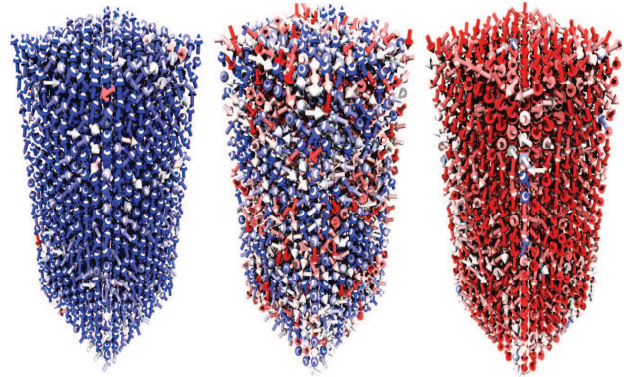}
\caption{\label{one} (color online) Snapshots of incoherent grain switching under the influence of the write field and thermal spot (see Fig. 1). The three images correspond to the beginning (-45 nm), middle (0 nm) and end (45 nm) of the field-temperature profile of Fig.1 appled at 25 nm/ns.  Blue white and red correspond to spins pointing up, sideways and down, respectively.}
\end{figure}

\begin{figure}[h]
\includegraphics[width=0.4\textwidth]{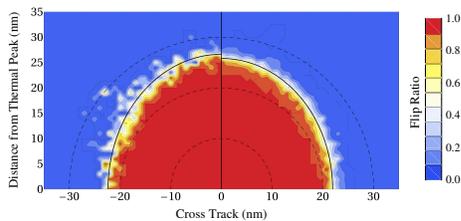}
\caption{\label{one} (color online) Contours of grain flip ratios from the application of the write field and thermal spot at distances from track center (horizontal) and delays/distances from peak temperatures (vertical). Left side shows results from atomic-level simulations ($M_s$ = 1000 emu/cc) and the right side using uniformly magnetized grain model from Ref. \onlinecite{torabi09}.}
\end{figure}  
  
   Fig. 2 shows the individual atomic-level spins as the grain moves down center track in the early case.  Switching is seen to be a highly localized and non-uniform process where the field acting on each spin individually serves to nucleate and drive the reversal process.  
   
   The effect of accounting for atomic-level spin interactions with the applied and stochastic fields is further illustrated by a comparison of results from the present method and that of Ref. \onlinecite{torabi09}, where temperature profiles $M(T)$ and $K(T)$ are integrated into a grain-level LLG simulation.  In our uniformly magnetized grain-level approach, $M(T)$ is calculated from atomic-level LLG simulations of the bulk material and $K(T)$ is determined from a power law $M^3$ dependence.\cite{torabi09} Fig. 3 shows the viable flip region contour resulting from the application of write field and thermal spot at various distances from track center and after the thermal peak. These results indicate that the region when a grain flip may occur is only slightly wider for the atomic-level simulation but that there is also considerably more fluctuations. In the case with $M_s$ = 400 emu/cc (corresponding to very large $H_K$ = 250 kOe), the write bubble is about 10\% larger for the atomic-level simulations (not shown) when compared with the uniform grain model. This larger `viable flip bubble' indicates that the atomic-level simulations can successfully transition to align with a reversed field at lower temperatures than the uniformly magnetized grain model, presumably due to the increased degrees of freedom.   

\begin{figure}[h]
\includegraphics[width=0.4\textwidth]{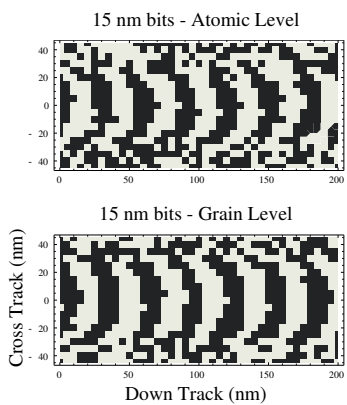}
\caption{\label{one} Recorded patterns of 15 nm bits comparing the atomic-level (top) and grain-level (bottom) calculations.}
\end{figure}
 
 \begin{figure}[h]
\includegraphics[width=0.5\textwidth]{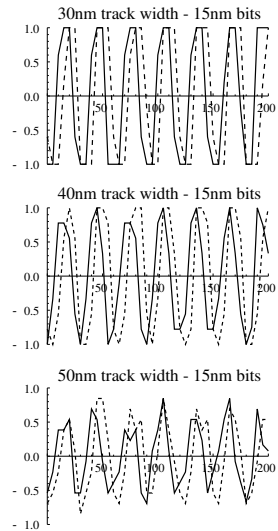}
\caption{\label{one} Magnetization profiles from the recorded patterns of Fig. 4 where continuous lines are from the atomic-level, and broken lines from the grain-level, simulations.  Averaging was performed over the center 15, 35 and 55 nm of the track width as indicated.}
\end{figure}  

   As a further test to compare the two models, simulations of recorded transitions with bit lengths of 5, 10, 15 and 20 nm were made by reversing the write field over a region of media 75 nm wide and 200 nm long. For the purposes of this preliminary comparison, magnetostatic interactions were again omitted. Transition patterns in the case of 15 nm long bits are shown in Fig. 4.  The atomic-level simulation results suggest a slightly worse transition profile.  This conclusion is further illustrated in Fig. 5 showing the results of track-width averaged magnetization profiles vs down-track direction of the 15 nm bit-length transitions in Fig. 4. Although this is not a conclusive test, these results, as well as those corresponding to other bit lengths (not shown), suggest a slightly smaller transition magnetization profile in the atomic-level case at the larger track widths, indicative of more pronounced fluctuations.
   
    In summary, micromagnetic simulations of recording media grain reversal at the atomic length scale are found to be important in order to capture thermal-fluctuation induced reversal mechanisms relevant to recording processes which occur at elevated temperatures. In contrast with simple models which mimic finite-temperature effects through the assignment of temperature profiles for the uniformly magnetized grain moment $M(T)$ and anisotropy $K(T)$, the present apporach accounts for individual atomic spin interactions with the applied write field through the Zeeman term  $-{\bf S}_i \cdot {\bf H}$.  In this way, atomic-level nucleation of grain reversal is included in the present model. Examination of an energy assisted recording  process reveals that the atomic-level simulations lead to more fluctuations in the viable write bubble, compared to the grain level model. The quality of the recorded transitions, as determined by cross-track averaged magnetization profiles in the down-track direction, appears worse in the atomic-level case due to these fluctuations. A significantly larger write bubble is found in the case of ultra-high $H_K$ = 250 kOe media. Proposed energy-assisted recording technology would benefit from the use of the atomic-level modeling approach to optimize design parameters and assess potential benefits. 
 
 This work was supported by Western Digital Corporation, the Natural
Science and Engineering Research Council of Canada (NSERC) and the Atlantic Computational Excellence network (ACEnet).

\end{document}